\documentclass[]{aa}

\usepackage{amsmath}
\usepackage[dvips]{graphicx}
\usepackage{natbib}
\bibpunct{(}{)}{;}{a}{}{,}
\usepackage[american]{babel}
\usepackage{txfonts}
\newcommand{\XSPECB}{\texttt{Xspec}}

\usepackage{xspace}

\def\ga{\mathrel{\mathpalette\fun >}}
\def\fun#1#2{\lower3.6pt\vbox{\baselineskip0pt\lineskip.9pt
  \ialign{$\mathsurround=0pt#1\hfil##\hfil$\crcr#2\crcr\sim\crcr}}}

{\catcode`\|=\active
  \gdef\Braket#1{\left<\mathcode`\|"8000\let|\bravert {#1}\right>}}
\def\bravert{\egroup\,\vrule\,\bgroup}

\def\gsim{\lower.5ex\hbox{\gtsima}}

\newcommand{\sax}{\textsl{BeppoSAX}\xspace}
\newcommand{\lecs}{\textsl{LECS}\xspace}
\newcommand{\mecs}{\textsl{MECS}\xspace}
\newcommand{\pds}{\textsl{PDS}\xspace}
\newcommand{\hp}{\textsl{HPGSPC}\xspace}

\newfont{\mc}{cmcsc10 scaled\magstep2}
\newfont{\cmc}{cmcsc10 scaled\magstep1}

\newcommand{\src}{4U~0115+63\xspace}

\newcommand{\bgi}{\begin{itemize}}
\newcommand{\edi}{\end{itemize}}

\newcommand{\be}{\begin{equation}}
\newcommand{\ee}{\end{equation}}

\newcommand{\bea}{\begin{eqnarray}}                  %
\newcommand{\eea}{\end{eqnarray}}                    %

\newcommand{\beaa}{\begin{eqnarray*}}                %
\newcommand{\eeaa}{\end{eqnarray*}}                  %

\newcommand{\bgd}{\begin{description}}
\newcommand{\edd}{\end{description}}

\newcommand{\bgf}{\begin{figure}}
\newcommand{\edf}{\end{figure}}

\newcommand{\bgc}{\begin{center}}
\newcommand{\edc}{\end{center}}

\newcommand{\bgt}{\begin{tabular}}
\newcommand{\edt}{\end{tabular}}

\newcommand{\bge}{\begin{enumerate}}
\newcommand{\ede}{\end{enumerate}}

\DeclareRobustCommand{\ion}[2]{%
 \relax\ifmmode
 \ifx\testbx\f@series
 {\mathbf{#1\,\mathsc{#2}}}\else
 {\mathrm{#1\,\mathsc{#2}}}\fi
 \else\textup{#1\,{\mdseries\textsc{#2}}}%
 \fi}

\def\jnl@style{\it}
\def\aaref@jnl#1{{\jnl@style#1}}

\def\aaref@jnl#1{{\jnl@style#1}}

\def\aj{\aaref@jnl{AJ}}                   
\def\araa{\aaref@jnl{ARA\&A}}             
\def\apj{\aaref@jnl{ApJ}}                 
\def\apjl{\aaref@jnl{ApJ}}                
\def\apjs{\aaref@jnl{ApJS}}               
\def\ao{\aaref@jnl{Appl.~Opt.}}           
\def\apss{\aaref@jnl{Ap\&SS}}             
\def\aap{\aaref@jnl{A\&A}}                
\def\aapr{\aaref@jnl{A\&A~Rev.}}          
\def\aaps{\aaref@jnl{A\&AS}}              
\def\azh{\aaref@jnl{AZh}}                 
\def\baas{\aaref@jnl{BAAS}}               
\def\jrasc{\aaref@jnl{JRASC}}             
\def\memras{\aaref@jnl{MmRAS}}            
\def\mnras{\aaref@jnl{MNRAS}}             
\def\pra{\aaref@jnl{Phys.~Rev.~A}}        
\def\prb{\aaref@jnl{Phys.~Rev.~B}}        
\def\prc{\aaref@jnl{Phys.~Rev.~C}}        
\def\prd{\aaref@jnl{Phys.~Rev.~D}}        
\def\pre{\aaref@jnl{Phys.~Rev.~E}}        
\def\prl{\aaref@jnl{Phys.~Rev.~Lett.}}    
\def\pasp{\aaref@jnl{PASP}}               
\def\pasj{\aaref@jnl{PASJ}}               
\def\qjras{\aaref@jnl{QJRAS}}             
\def\skytel{\aaref@jnl{S\&T}}             
\def\solphys{\aaref@jnl{Sol.~Phys.}}      
\def\sovast{\aaref@jnl{Soviet~Ast.}}      
\def\ssr{\aaref@jnl{Space~Sci.~Rev.}}     
\def\zap{\aaref@jnl{ZAp}}                 
\def\nat{\aaref@jnl{Nature}}              
\def\iaucirc{\aaref@jnl{IAU~Circ.}}       
\def\aplett{\aaref@jnl{Astrophys.~Lett.}} 
\def\apspr{\aaref@jnl{Astrophys.~Space~Phys.~Res.}}
\def\bain{\aaref@jnl{Bull.~Astron.~Inst.~Netherlands}} 
\def\fcp{\aaref@jnl{Fund.~Cosmic~Phys.}}  
\def\gca{\aaref@jnl{Geochim.~Cosmochim.~Acta}}   
\def\grl{\aaref@jnl{Geophys.~Res.~Lett.}} 
\def\jcp{\aaref@jnl{J.~Chem.~Phys.}}      
\def\jgr{\aaref@jnl{J.~Geophys.~Res.}}    
\def\jqsrt{\aaref@jnl{J.~Quant.~Spec.~Radiat.~Transf.}}
\def\memsai{\aaref@jnl{Mem.~Soc.~Astron.~Italiana}}
\def\nphysa{\aaref@jnl{Nucl.~Phys.~A}}   
\def\physrep{\aaref@jnl{Phys.~Rep.}}   
\def\physscr{\aaref@jnl{Phys.~Scr}}   
\def\planss{\aaref@jnl{Planet.~Space~Sci.}}   
\def\procspie{\aaref@jnl{Proc.~SPIE}}   

\begin{document}

\title{Study of the accreting pulsar \src with a bulk and thermal
Comptonization model}

\author{Carlo Ferrigno
        \inst{1,2}
	\and
      	Peter A. Becker\inst{3}
        \and
        Alberto Segreto\inst{4}
        \and
        Teresa Mineo\inst{4}
        \and
        Andrea Santangelo\inst{2}
}

\authorrunning{C. Ferrigno et al}
\titlerunning{A Comptonization model for 4U~0115+63}
   \offprints{C. Ferrigno}

\institute{ISDC chemin d'\'Ecogia, 16 1290 Versoix Switzerland\\
	\email{Carlo.Ferrigno@unige.ch}
          \and
           IAAT, Abt.\ Astronomie, Universit\"at T\"ubingen,
           Sand 1, 72076 T\"ubingen, Germany
           \and
           Department of Computational and Data Sciences \\George Mason University\\
	   4400 University Drive, MS 6A3 Fairfax, VA 22030
          \and
          IASF-INAF, via Ugo la Malfa 153, 90146 Palermo Italy        
          }

\date{Received ---; accepted ---}

\abstract
	{Highly magnetized pulsars accreting matter in a binary system are bright
         sources in the X-ray band (0.1--100\,keV). Despite the early comprehension
         of the basic emission mechanism, their spectral energy distribution is generally
         described by phenomenological or simplified models.} 
	{We propose a study of the spectral emission from the high mass
         X--ray binary pulsar \src by means of thermal and bulk Comptonization models
         based on the physical properties of such objects.} 
        {For this purpose, we analyze the \sax data in the energy range 0.7--100\,keV  
        of the 1999 giant outburst, 12 days after the maximum.
	We focus first on the phase averaged emission, and then on the phase resolved
        spectra, by modeling the system using a two-component continuum.} 
        {At higher energy, above $\sim 7$\,keV, the emission is due to thermal and
         bulk Comptonization of the seed photons produced by cyclotron cooling of
         the accretion column, and at lower energy, the emission is due to thermal
         Comptonization of a blackbody source in a diffuse halo close to the stellar
         surface. Phase resolved analysis establishes that most of the emission in
         the main peak comes from the column, while the low energy component gives
         a nearly constant contribution throughout the phase.}  
	{From the best fit parameters, we argue that the cyclotron emission
         is produced $\sim 1.7$\,km above the stellar surface, and escapes from
         the column near its base, where the absorption features are generated by
         the interaction with the magnetic field in the halo. We find that in \src,
         the observed spectrum is dominated by reprocessed cyclotron radiation, whereas
         in other bright sources with stronger magnetic fields such as Her~X-1, the spectrum
         is dominated by reprocessed bremsstrahlung.} 

   \keywords{X-rays: binaries, pulsars: individual: 4U 0115+63}

\maketitle

\section{Introduction}

X--ray binary pulsars (XRBPs) were discovered more than 35 years ago
with the pioneering observations of the bright sources Her~X-1
\citep{giac71} and Cen~X-3 \citep{tananbaum1972}. The basic mechanisms
of the pulsed emission were understood quickly: radiation originates
from the accretion of ionized gas onto a rotating neutron star (NS) from
a nearby companion, an O or B star ($M \geq 5M_{\odot}$) in high mass
X-ray binaries (HMXBs), a later than type A star ($M\leq M_{\odot}$) in
low mass X-ray binaries. For strong magnetic fields
($B\sim10^{11-12}$\,G) of the compact object, the plasma is threaded at
several hundred stellar radii and then funneled along the magnetic field
lines onto the magnetic poles, forming one or two accretion columns
\citep{Pringle72,Davids73}. The system is powered by the conversion of
gravitational potential energy into kinetic energy, which is eventually
emitted in the form of X-rays as the plasma decelerates, possibly
through a radiative shock, and settles onto the stellar surface.
Pulsation is generated because of the nonalignment between the magnetic
and rotational axes.

The spectral emission from XRBPs still presents many puzzling aspects.
Recently, \citet{becker2007} made a major step forward by proposing a
model (from here on the BW model) which represents a complete
calculation of the X-ray spectrum associated with the physical accretion
scenario first suggested by \citet{Davids73}. A summary of the model
properties will be given in Sect.~\ref{sec:bwmod}

A unique spectral characteristic of many HMXBs is the presence of
cyclotron resonant scattering features (CRSFs), which provide a tool
for direct measurement of the magnetic-field strengths of accreting
pulsars, a key ingredient of the model. The fundamental line appears at
$E_{\rm cyc} = 11.6\,B_{12}\times (1+z)^{-1}$\,keV, where $B_{12}$ is the
magnetic field strength in units of $10^{12}$\,G and $z$ is the
gravitational redshift in the line-forming region \citep{wasserman1983}.
However, the determination of the centroid energy of the fundamental
line might be hampered by its complex shape, due to photon spawning
\citep[see e.g.][]{ingo2005}. The second harmonic is deeper and can be
described by a simple analytical function such as a Gaussian curve,
since it is due to pure absorption \citep{nishimura2003}. Phase resolved
spectral analysis is essential for this kind of study because the
spectral characteristics of the lines depend heavily on the angle
between the observation direction and the magnetic field orientation
\citep[see e.g. ][]{araya1999,araya2000,gabi2007}.

In this paper we use the BW model to study an observation of the
bright HMXB 4U~0115+634. Hard X-ray radiation from this source was discovered
in 1969 \citep{whitlock1989}, during one of its giant type II outbursts
(16 have been observed up to now), when the X-ray luminosity rises to a
few $10^{37}$\,erg/s, two orders of magnitude above its quiescent level.
X-ray pulsations ($P_S=3.6$\,s) were soon discovered
\citep{cominski1978}, and the determination of the orbital parameter
($P_\mathrm{orb}=24.3\,\mathrm{d}$, $e=0.34$, $a_X \sin i =
140.1\,\mathrm{lt-s}$) was made possible for the first time using the
\textsl{SAS} data \citep{rappaport1978}. The high energy ($\ga 10$\,keV)
continuum of \src has previously been modeled with the standard
power-law modified by an exponential cut-off \citep[e.g.][]{coburn2002}.
The source has also been widely studied in the optical and IR bands,
allowing the identification of its optical counterpart, V~635~Cas
\citep{johns1978}, and also the determination of its distance 7--8\,kpc
\citep{negueruela2001a}.

In the spectrum of \src, cyclotron harmonics have been observed up the
fifth order \citep{wheaton1979,white1983,heindl1999,santangelo1999},
while in the other known XRBPs, only the second harmonic, and in one
case the third \citep{tsygankov2006}, is sometimes detected
\citep{coburn2002, orlandini2004,isabel2007}. The luminosity of \src
during outburst is comparable to that of the bright sources Her~X-1 and
Cen~X-3 studied by BW, but the magnetic field strength, $\sim
10^{12}$\,G, implied by the cyclotron absorption features in the
spectrum of \src is considerably lower than the field strength in the
other two sources. As a result of the analysis presented here, we find
that the decrease in the field strength causes a fundamental shift in
the character of the emitted X-ray spectrum, from Comptonized
bremsstrahlung in Her~X-1 and Cen~X-3 to Comptonized cyclotron in
\src.

The relation between the optical properties of the system and the X-ray
type II outbursts have been studied by \cite{whitlock1989}, who
suggested a 3-year quasi-periodicity of the outbursts. Later,
\citet{negueruela2001a} and \citet{negueruela2001b} were able to
characterize the outburst mechanism as a result of a disc
instability due to radiative warping.

The \textsl{GINGA} observations evidenced how both the continuum and the
line energies are phase-dependent \citep{mihara2004}. A remarkable
result, obtained analyzing the \textsl{RXTE} data, is the luminosity
dependence of the cyclotron features \citep{nakajima2006}. This is
interpreted as reflecting the passage from a radiation-dominated flow
with a shock above the stellar surface at high luminosity, to a
matter-dominated flow at low luminosity, in which the accreting material
is brought to rest at the stellar surface by the pressure of the gas.
Additional constraints on the height of the line forming region in \src
have been developed by \citet{russi2007}, who also concluded that the
height of the hard-photon emission region in the column decreases with
decreasing luminosity.

After providing a brief summary of the \citet{becker2007} column
emission model, in Sect.~\ref{sec:observation} we describe the
observations and the analysis methods. In Sect.~\ref{sec:results} we
report our results, which are discussed in detail in Sect.~\ref{sec:discussion}.
Finally, in Sect.~\ref{sec:conclusions}, we draw our conclusions.

\section{Summary of the column emission model}
\label{sec:bwmod}
\citet{becker2007} proposed a model for the production of the accretion
column emission spectrum based on the bulk and thermal Comptonization of
seed photons emitted via bremsstrahlung, cyclotron, and blackbody
processes occurring in the accreting plasma. The partial differential
equation governing the Green's function yields an analytical solution
that can be convolved with the seed photon distribution to produce an
emergent spectrum displaying a power-law continuum with an exponential
cutoff at high energies. They assume a cylindrical geometry for the
column, along with a constant electron temperature and a constant
magnetic field. The constant properties of the column are a reasonable
approximation, since most of the relevant interactions are thought
to take place in a relatively thin layer close to its base. The density
and velocity profiles of the infalling matter are analytical
approximations of numerically computed functions, and the
angle-dependent cross section is approximated using two terms to
represent the mode-averaged electron scattering cross sections for
photons propagating either parallel or perpendicular to the magnetic
field direction, denoted by $(\sigma_\parallel)$ and $(\sigma_\perp)$,
respectively.

The cyclotron and bremsstrahlung source distributions each vary as the
square of the plasma density, but their energy dependences are quite
different: the cyclotron emission is due to the collisional excitation
and radiative decay of the first Landau level, and it therefore displays
a monoenergetic energy dependence; the bremsstrahlung emission is due to
the braking of electrons interacting with protons, and it therefore
possesses a continuous energy dependence. The combination of these terms
provides a good approximation of the free-free process in very strong
magnetic fields \citep{riffert1999}. It should be emphasized that when
the cyclotron energy is comparable to the plasma temperature, the
cyclotron emission becomes the preferred cooling channel
\citep{arons1987}. Finally, blackbody emission is localized at the base
of the column due to the presence of the thermal mound; it is the only
term for which an integral must be numerically computed to derive the
emergent spectrum.

For given values of the stellar mass $M_*$ and the stellar radius
$R_*$, the BW model has six fundamental free parameters, namely the
accretion rate $\dot M$, the column radius $r_0$, the electron
temperature $T_e$, the magnetic field strength $B$, the photon diffusion
parameter $\xi$, and the Comptonization parameter $\delta$, with the final
two parameters are defined by
\begin{equation}
\xi=\frac{\pi r_0 m_p c}{\dot M \left(\sigma_\parallel
\sigma_\perp\right)^{1/2}} \ , \ \ \ \ 
\frac{\delta}{4}=\frac{y_\mathrm{bulk}}{y_\mathrm{thermal}} \ ,
\label{eq:firsteq}
\end{equation}
where $m_p$ denotes the proton mass, $c$ is the speed of light,
and $y_\mathrm{bulk}$ and $y_\mathrm{thermal}$ represent the
$y$-parameters\footnote{The $y$-parameters describes the average
fractional energy change experienced by a photon before it escapes
through the column walls.} for the bulk and thermal Comptonization
processes.

The accretion rate is constrained by the observed luminosity, under the
assumption of isotropic accretion and unitary efficiency, which might allow
us to remove this parameter from the set. A unique solution for the
remaining free parameters can be found by comparing the model spectrum
with the observations for a particular source. The analytical nature of
the model makes it amenable to incorporation into standard analysis
packages such as \XSPECB, as we did, which facilitates observational
testing.

\section{Observations and data analysis.}
\label{sec:observation}
This work is based on the observations performed by \sax \citep[see ][
for a description of the mission]{sax} on 1999 March, 26 at 17:31:32.5
(OP6714) and ended on 1999 March, 27 at 17:34:05.5, during the decay of
a giant outburst, more precisely 12 days after the peak, when the flux
in the 2--10\,keV energy range was a factor of two lower than the
maximum value \citep[see Fig.~1 of ][]{heindl1999}. Among the many
publicly available observations of \src during its X-ray outbursts, our
choice is driven by the wide energy band, good spectral resolution, and
reliable calibration of the \sax instruments, and also by the
relatively long exposure of this data-set. Another observation of the
same outburst, made seven days earlier, is reported in
\citet{santangelo1999}.

We analyze the data obtained by the high energy narrow field instruments
\lecs \citep[0.7--4.0\,keV, ][]{lecs}, \mecs \citep[1.5--10.5\,keV,
][]{mecs}, \hp \citep[7--44\,keV, ][]{hp} and \pds \citep[15--100\,keV,
][]{pds} after the processing through the standard pipeline
(\texttt{saxdas} v.2.1). Background subtraction is performed using the
Earth occultation for \hp, the off-source pointing for \pds, and the
standard calibration files for \mecs and \lecs. The \textsl{LECS} and
\textsl{MECS} source extraction radii are set to $8^\prime$ for an
optimal signal to noise ratio. The instrument exposure times are 5043\,s
(\lecs), 53~660.4\,s (\mecs), 42~495.3\,s (\hp), and 48~328.9\,s (\pds),
with the differences due to the variations in the operating
procedures. 

\subsection{Spectral analysis} The spectral analysis is performed with
the \XSPECB~package version 12.3.1 \citep{xspec}, using our implementation of
the analytical model for the vertically-integrated accretion column
emission spectrum developed by \citet{becker2007}. The fundamental
and harmonic cyclotron resonant absorption features are modeled using
multiplicative Gaussian functions:
\begin{equation}
G(E) = \left(1 - \frac{N_n}{\sigma_n\sqrt{2\pi}}e^{-\frac{1}{2}{\left(\frac{E-
E_{cn}}{\sigma_n}\right)}^2}\right) \ ,
\label{eq:gaus}
\end{equation}
where $N_n$, $E_{cn}$, and $\sigma_n$ denote the strength, energy, and
standard deviation of the absorption function for the $n$th harmonic.
The line width is fixed whenever it cannot be determined by the fit;
this is always the case for the fourth and fifth harmonics, set to the
convenient value of 4\,keV. 

The phase averaged spectrum is studied at first, then we divide the
pulse period into 15 bins of equal width and analyze each spectrum
separately to obtain the dependency of the parameters on the phase.
Due to the short exposure, we use the \lecs data only in the phase
averaged spectrum. All of the errors are reported at 90\% confidence
level (c.l.), i.e. $\Delta\chi_\mathrm{min}^2=2.7$ for one parameter of
interest, unless stated differently. 

\subsection{Timing analysis}
We transform the photon arrival times into the Solar system barycenter
reference frame using the standard tool for \sax (\texttt{baryconv}),
then to the binary system reference frame using the ephemeris
\citep{rappaport1978,tamura1992}. Using the \hp data, we find the pulse
period $P=3.614246\pm0.000002$\,s on 1999-03-26 at 17:28:45.643\,UT
(MJD~51263.72830605) adopting the phase shift method described by
\cite{ferrigno2007}. I.e. we divide the observation into 174 equally
spaced time bins; in each bin, we determine the phase of the pulse
maximum with a typical uncertainty of 5\% and then linearly fit these
values as functions of time by adapting iteratively the initial
guess on the period. The quoted uncertainty is purely statistical. If no
orbital correction is applied, we find a significant variation of the
pulse period within the observation: $\dot P = (2.6904 \pm
0.0005)\times10^{-9}$\,s/s.

We fold the light curves in different energy ranges with 100 phase bins
(Fig.~\ref{fig:pulses}). To account for the timing systematics of the
three instruments the pulse profiles are arbitrarily shifted to obtain
an optimal alignment in the common energy ranges (8--11\,keV for \mecs
and \hp, 20--25\,keV for \hp and \pds): we apply a forward phase shift
of 4 bins ($\sim$0.14\,s) to the \mecs profiles and of 5 bins
($\sim$0.18\,s) to the \pds ones keeping the \hp profiles as reference.
The reason for this instrumental timing problem is currently unknown.

\section{Results}
\label{sec:results}
\subsection{Pulse profiles}

\begin{figure}
  \begin{center}
  \resizebox{\hsize}{!}{
      \includegraphics[angle=0]{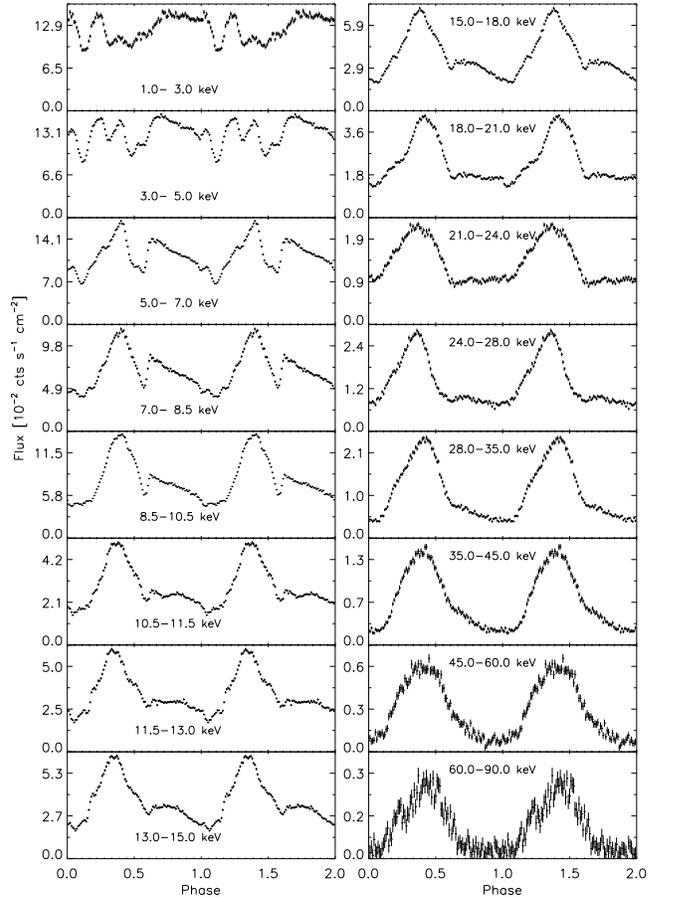}
	}
	\caption{Pulse profiles of 4U~0115+63 in different energy ranges. 
		The data are from the
		\mecs (1.0--8.5\,keV), \hp (8.5--28\,keV) and \pds (28--95\,keV) instruments on board of \sax.
    }
\label{fig:pulses}
\end{center}
\end{figure}

The shape of the pulse profiles, shown in Fig.~\ref{fig:pulses}, is
highly variable with energy: above $\sim10$\,keV we can clearly
distinguish a main peak at pulse phase $\sim0.4$, and, with more
attention, a secondary shallower one at pulse phase $\sim0.7$
interpreted by some authors as the partially obscured emission of the
column of the opposite polar cap \citep{russi2007}. Below this energy,
the pulse shape is more complex, suggesting that the lower energy
emission is generated by a different mechanism.

\begin{figure}
  \begin{center}
      \resizebox{\hsize}{!}{\includegraphics[angle=0]{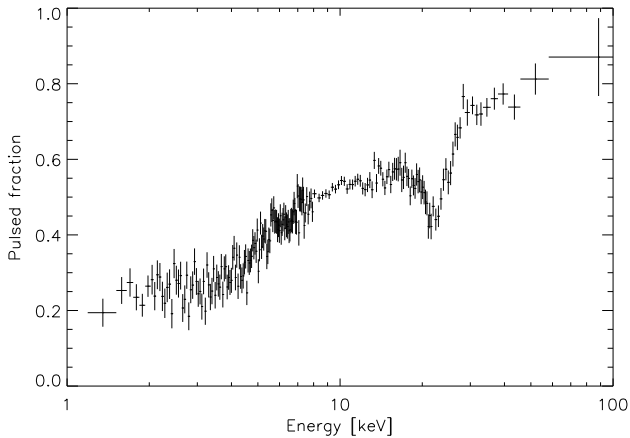}}
	\caption{Pulsed fraction as function of energy, from 1.2 to 8\,keV the data are from MECS, from 8 to 30\,keV from \hp, and
	from 30 to 120\,keV from \pds.}
\label{fig:pulsed_fraction}
\end{center}
\end{figure}

We compute the pulsed fraction, i.e. the ratio $\frac{F_\mathrm{max} -
F_\mathrm{min}}{F_\mathrm{max}+ F_\mathrm{min}}$, as a function of
energy, where the count rates at the maximum ($F_\mathrm{max}$) and
minimum ($F_\mathrm{min}$) of the pulses are the weighted average of
three neighboring phase bins. We perform an energy binning, which
ensures that the signal is at least 140 times larger than the noise for
each pulse profile.

As shown in Fig.~\ref{fig:pulsed_fraction}, the pulsed fraction below
5\,keV remains always between 20\% and 30\%, then it begins to raise,
showing at $\sim22$\,keV a sharp localized decrease due to the resonant
scattering at the second Landau level. The low S/N at high energy
prevents the detection of the higher harmonics, and even the fundamental
cannot be clearly identified despite the high count level. This may be
due to competing effects such as photon spawning and cyclotron
emission. We verified that in the overlapping energy ranges, the data
of the three instruments are consistent with each other.

\subsection{Phase averaged spectrum}

Due to the well-known problems in the absolute calibration, we used
inter-calibration constants between the different instruments: they are
obtained from a best fit procedure setting the \hp as reference. We
verified that the results reported in Table \ref{tab:constants} are
model independent (within the quoted errors) and consistent with the instrument team
recommendations.\footnote{ \scriptsize \texttt{http://heasarc.gsfc.nasa.gov/docs/sax/abc/saxabc/saxabc.html}}

\begin{table}
\caption{Instrument inter-calibration constants using the \hp as reference.}
\begin{center}
 \begin{tabular}{ l r@{}l }
\hline
\hline
Instrument         & \multicolumn{2}{c}{Constant} \\
\hline
\lecs	  & 0.814  & $\pm$ 0.008 \\
\mecs  & 0.970  & $\pm$ 0.003 \\
\pds     & 0.857  & $\pm$ 0.004 \\
\hline
\end{tabular}
\label{tab:constants}
\end{center}
\end{table}

\subsubsection{Phenomenological model}
The aim of this work is the introduction of a physical model to describe
the spectrum. However, we study the traditional phenomenological models
for comparison with the results present in the literature. The fit
with a simple high-energy cut-off power law plus absorption features is
unsatisfactory ($\chi^2_\mathrm{red}=2.4$); we then adopt the model that
\citet{dima2008} applied to the high mass X-ray binary EXO~2030+375: a
broad Gaussian emission line plus a high-energy cut-off power-law.
The absorption lines are described by eq.~(\ref{eq:gaus}), photoelectric
absorption by the model \texttt{phabs} in \XSPECB.

\begin{figure}
  \begin{center}
      \resizebox{\hsize}{!}{\includegraphics[angle=0]{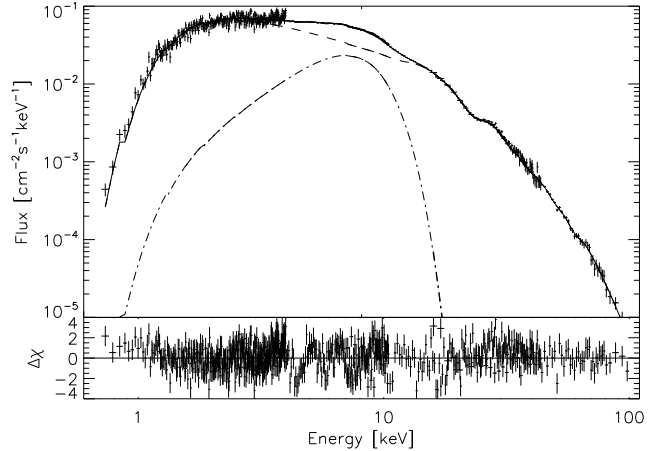}}
	\caption{Upper panel: the unfolded spectrum of \src described by the phenomenological model: the dashed line is the 
		high-energy cut-off power law, the dot-dashed line the broad Gaussian, and the solid line the sum. The spectra of the four instruments have been scaled according to the inter-calibration constants of Table~\ref{tab:constants}. Lower panel: residuals from the best fit model.}
\label{fig:phase_averaged_pheno}
\end{center}
\end{figure}

The unfolded spectrum is reported in Fig.~\ref{fig:phase_averaged_pheno}
and the best fit parameters in Table~\ref{tab:phase_averaged}: the
continuum is altered only locally by cyclotron resonances, the
fundamental line is shallower than the second harmonic as expected from
the pulsed fraction of Fig.~\ref{fig:pulsed_fraction}, and the photon
index is almost equal to one, as typical for saturated Comptonization
\citep{coburn2002}. The photoelectric absorption is parametrized by
$N_\mathrm{H} =(1.38\pm0.03)\times 10^{22}\,\mathrm{cm^{-2}}$, which is
higher than the value for Galactic absorption in the source direction,
$N_\mathrm{H,Gal} =0.9\times 10^{22}\,\mathrm{cm^{-2}}$, \citep[average
of the values of ][]{dickey1990,kalberla2005}, thus the X-ray emission
seems to be attenuated also locally around the source.

We derive a flux in the 1-100\,keV band of $(1.69\pm0.02) \,
\mathrm{erg\,s^{-1}\,cm^{-2}}$ corresponding to a luminosity
$L_X=6.3\times10^{37}\,\mathrm{erg\,s^{-1}}$ assuming a distance of
7\,kpc. In this model, the power-law component is responsible for the
high- and low-energy portions of the spectrum, which are characterized
by completely different pulse profiles. The ``bump'' below the cyclotron
energy is described by the broad Gaussian line, which does not
correspond clearly to any physical process.

\subsubsection{Comptonization model}
If we apply the model proposed by \citet{becker2007}, using as an
input parameter the magnetic field derived using the cyclotron
absorption features, then the Comptonized cyclotron emission has a flux
about two orders of magnitude larger than the bremsstrahlung and
blackbody components, and the resulting spectrum does not reproduce the
lower energy portion of the \src spectrum. Driven by the observation
that the pulse profiles below $\sim5$\,keV are very different from the
others, we therefore try to fit the spectrum below this energy with
another component. The attempts to use a blackbody, a cut-off power law,
or a simple power-law, all attenuated by photoelectric absorption, are
unsuccessful, giving $\chi_\mathrm{red}\ga2.5$. Among the reasonably
simple models, only the thermal Comptonization of low energy
(0.1--0.5\,keV) blackbody emission, located in the center of a spherical
cloud of hot electrons, gives a satisfactory fit to the data, as
verified by the application of the \texttt{compST} \citep{compst} and
\texttt{compTT} \citep{comptt} models in \XSPECB. We will
therefore attempt to fit the broad band spectrum of \src using a
combination of the BW bulk+thermal Comptonization model, which gives
good agreement at high energies, and a thermal Comptonization component
at lower energy. As discussed below, we interpret the low energy
component as emission arising in a halo of gas surrounding the base of
the accretion column.

From the statistical point of view, it is not possible to discriminate
between the thermal Comptonization models for the low energy component.
We then compute the effective radius of the blackbody which produces the
seed photons following the method of \citet{jean1999}. In the
\texttt{compST} model, the temperature of the seed photons is fixed to
0.1\,keV: for best fit parameters of the hot electron temperature $T_e=
(2.91\pm0.04)$\,keV, the optical depth of the spherical cloud
$\tau=22.2\pm0.4$, and the 2-100\,keV flux for this spectral component
of $5.13\times 10^{-9}\,\mathrm{erg\,cm^{-2}s^{-1}}$, the blackbody
radius is 550\,km, which is clearly too high for such a system. We
therefore utilize the \texttt{compTT} model rather than the
\texttt{compST} model to analyze the low-energy component.

Using the \texttt{compTT} model, it is possible to determine the seed
photon temperature, $T_0=(0.51\pm0.01)$\,keV, if we fix the absorption
column to the Galactic value. In this scenario, the low energy emission
is produced by an optically thick plasma halo ($\tau=24.4\pm0.4$) at
temperature $T_C=(2.90\pm0.04)$\,keV, while for comparison the BW
electron temperature in the column is $T_e=(7.94\pm0.11)$\,keV. With these
parameters (reported in Table~\ref{tab:phase_averaged}), corresponding
to a flux in the 0.5--100\,keV band of $5.19\times 10^{-9} \,
\mathrm{erg\,cm^{-2}s^{-1}}$, we obtain a blackbody effective radius of
$\sim$16\,km, which is of the right order of magnitude. Assuming a
standard NS, the gravitational redshift at the surface is 0.3.
Correcting for this effect, we find best-fit parameters corresponding to
a blackbody with $\sim$9\,km radius. However, in the rest of the paper
we will neglect the relativistic effects for consistency with the BW
model.

If we fix the magnetic field strength of the BW component using the
value derived from the absorption lines, then we find there is no
combination of the other parameters which can account for the emission
between $\sim7$ and $\sim10$\,keV. We therefore leave the magnetic field
in the BW component as a free parameter, which effectively separates the
spatial region where the high-energy BW component is produced from the
region where the cyclotron absorption features are imprinted on the
emergent spectrum. The implications of this physical picture are further
discussed below. The mass accretion $\dot M$ is also determined through
the spectral fit. Since there is a very strong correlation between the
column radius $r_0$ and $\dot M$, we study the contour plot of $r_0$ and
$\dot M$ and find that a good fit is obtained for $r_0\simeq 600$\,m and
$\dot M \simeq 0.6 \times 10^{17}$\,g/s. The value obtained for the
accretion rate using the fit is somewhat lower than that expected if the
emission were isotropic, in which case the X-ray luminosity would imply
$\dot M = 3.3 \times 10^{17}$\,g/s. Our result for $\dot M$ therefore
suggests that the emission from the column is anisotropic, as expected
for a fan beam. This point is further discussed in Sect.~\ref{sec:discussion}.

\begin{figure}
  \begin{center}
      \resizebox{\hsize}{!}{\includegraphics[angle=0]{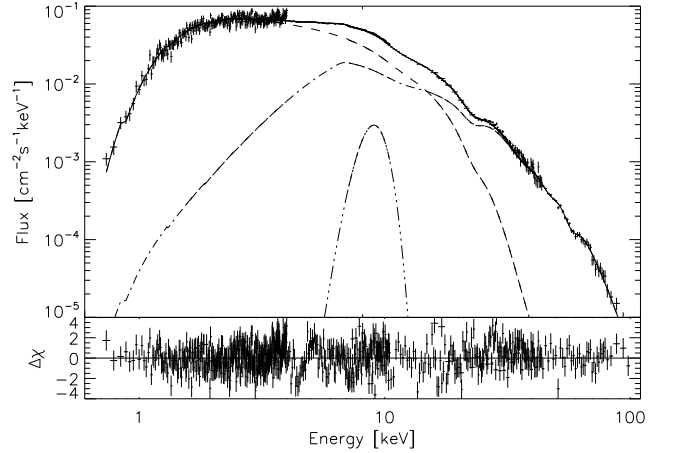}}
	\caption{Upper panel: the unfolded spectrum of \src described by the Comptonization model: the dashed line is the 
		thermal Comptonization, the dot-dashed line the column emission computed using the BW model, the dot-dot-dashed line a
		narrow Gaussian line, and the solid line the sum. The spectra of the four instruments have been scaled according to the inter-calibration constants of Table~\ref{tab:constants}. Lower panel: residuals from the best fit model.}
\label{fig:phase_averaged}
\end{center}
\end{figure}

We then freeze $r_0$ and $\dot M$ to obtain a stable fit for the other
model parameters, and the neutron star mass and radius are fixed at the
standard values of $1.4\,M_\odot$ and 10\,km, which cannot be
constrained by the comparison to the data. Among the results reported in
Table~\ref{tab:phase_averaged}, we note that the magnetic field which
produces the cyclotron emission in the BW model has a strength
$B=(5.91\pm0.02)\times 10^{11}$\,G while the line centroid energy
implies that the magnetic field in the line forming region is $\sim
10^{12}$\,G. These results clearly suggest the existence of a spatial
offset between the BW emission region and the cyclotron absorption
region, as further explored in Sect.~\ref{sec:discussion}.

Due to some residuals in the transition region between the low energy
thermal Comptonization component and the BW column emission, we
introduce a third emission component in the form of a Gaussian line with
energy $\sim 9\,$keV and standard deviation $\sigma=1$\,keV. From
Fig.~\ref{fig:phase_averaged} it is clear that this component gives a
negligible contribution to the flux, unlike in the phenomenological
model, and its necessity here probably stems from our rough description of
the cyclotron scattering in strong magnetic fields.

\begin{table*}
\caption{Best fit parameters of the phase averaged spectral model, if
uncertainties are omitted, the parameter is frozen.}
\begin{center}
 \begin{tabular}{ l r@{}l | l r@{}l}
\hline
\hline
\multicolumn{3}{c}{Phenomenological} & \multicolumn{3}{c}{Physical}\\
 \hline
$N_\mathrm{H}$ [$\mathrm{cm^{-2}}$]    &            1.38 &$\pm0.03$ & $N_\mathrm{H}$ [$\mathrm{cm^{-2}}$]    &               0.9 &              \\
\hline
				&	&  & $T_0$ [keV]			&   $0.51$  & $\pm 0.01$ \\
				&   	&  & $T_C$ [keV]			&   $2.90$ & $_{-0.05}^{+0.03}$\\
				& 	&  & $\tau$				&   $24.4$ & $\pm 0.4$\\
				&      &  & norm$^*$				&   $0.257$ & $\pm 0.003$\\
\hline                                                                         
$E_c$ [keV] & 14.5 & $\pm 0.3$    &     $\xi$                                  & $4.0$ & $_{-0.4}^{+0.2}$\\
$E_f$ [keV] & 13.5 & $\pm 0.2$    &     $\delta$                               & $0.59$ & $\pm 0.02$\\
$\Gamma$    & 0.998& $\pm 0.007$  &     $B\,[\mathrm{10^{12} G}]$              & $0.591$ & $\pm 0.002$\\
$K$         & 0.247& $\pm 0.003$  &     $T_e$ [keV]                            & $7.94$ & $\pm 0.11$ \\
 & & & $r_0$ [m]                              & 600                &                         \\
 & & & $\dot M\,[\mathrm{10^{17} g\,s^{-1}}]$ & 0.6                &                         \\
\hline                                                                         
$E_G$ [keV]                          & $7.056$ & $_{-0.029}^{0.014}$ & $E_G$ [keV]                           &  $9.07$ & $\pm 0.12$\\
$\sigma_G$ [keV]                     & 2.51 &$_{-0.03}^{+0.08}$  &$\sigma_G$ [keV]               & 1 & \\
$N_G$ $[\mathrm{cts\,cm^{-2}\,s^{-1}}]$  & 0.144 & $\pm0.003$ & $N_G$ $[\mathrm{cts\,cm^{-2}\,s^{-1}}]$  & $0.0080$ & $\pm 0.0009$\\
\hline                                                                         
$E_1$ [keV]                          & 11.16 & $\pm0.15$ & $E_1$ [keV]                          & $11.58$ & $\pm 0.13$\\
$\sigma_1$ [keV]                     & 1 & & $\sigma_1$ [keV]                     & 2 & \\
$N_1$ $[\mathrm{cts\,keV}]$  & 0.09 & $\pm0.02$ & $N_1$ $[\mathrm{cts\,keV}]$  & 0.70 & $\pm0.04$ \\
\hline                                                                         
$E_2$ [keV]                          & 22.99 & $\pm0.05$ & $E_2$ [keV]                          & $22.76$ & $\pm 0.09$\\
$\sigma_2$ [keV]                     & 2.8 &$\pm0.1$ &$\sigma_2$ [keV]                     & 2 & \\
$N_2$ $[\mathrm{cts\,keV}]$  & 2.3 & $\pm0.1$ & $N_2$ $[\mathrm{cts\,keV}]$  & $1.22$ & $\pm0.05$ \\
\hline                                                                         
$E_3$ [keV]                          & 32.6 & $\pm0.4$ &$E_3$ [keV]                          & $36.3$ & $\pm 0.6$\\
$\sigma_3$ [keV]                     & 3 &  &$\sigma_3$ [keV]                     & 3 &  \\
$N_3$ $[\mathrm{cts\,keV}]$  & 1.6 & $\pm0.2$&$N_3$ $[\mathrm{cts\,keV}]$  & $2.4$ & $\pm 0.2$\\
\hline                                                                         
$E_4$ [keV]                          & 40.8 & $\pm0.7$ &$E_4$ [keV]                          &  $46.5$ & $\pm 1.1$ \\
$\sigma_4$ [keV]                     & 4 &  &$\sigma_3$ [keV]                     & 4 &  \\
$N_4$ $[\mathrm{cts\,keV}]$  & 2.5  & $\pm0.2$&$N_4$ $[\mathrm{cts\,keV}]$  & $2.9$ & $\pm 0.4$\\
\hline                                                                         
$E_5$ [keV]                          & 53 & $\pm1$ &$E_5$ [keV]                          & $57$ & $\pm 2$\\
$\sigma_5$ [keV]                     & 4 &  &$\sigma_3$ [keV]                     & 4 &  \\
$N_5$ $[\mathrm{cts\,keV}]$  & 2.6 & $\pm0.4$&$N_5$ $[\mathrm{cts\,keV}]$  & $3.6$ & $\pm 0.7$\\
\hline
$\chi^2$ (d.o.f.)                    & 1.48 &(705) &$\chi^2$ (d.o.f.)                    & 1.49 &(703) \\
flux$^{**}$                          &$ (1.070$ & $\pm0.002 )\times 10^{-8}$ &flux$^{**}$  & $ (1.070$ & $\pm0.003 )\times 10^{-8}$ \\
\hline
\end{tabular}
\\$^{*}$ the definition of the model normalization is reported in \XSPECB~documentation.
\\$^{**}$ the flux in the 1-100\,keV energy band is expressed in units of $\mathrm{erg \, cm^{-2}\, s^{-1}}$.
\label{tab:phase_averaged}
\end{center}
\end{table*}

\subsection{Phase resolved spectroscopy}

The vertically-integrated accretion column emission spectrum computed
using the BW model is independent of the observation angle of the
observer, who is assumed to be at rest in the frame of the neutron star.
The angle independence stems from the fact that the model does not
incorporate the full angle and energy dependence of the cyclotron
scattering cross section, but instead uses a two-component formalism
based on the cross sections $(\sigma_\parallel)$ and $(\sigma_\perp)$
for photons propagating either parallel or perpendicular to the field.
Despite the simplified treatment of the scattering cross section, it is
interesting to see what insight can be gained by fitting the BW model to
the phase-resolved spectral data. Physically, variations in the
phase-resolved spectrum arise due to the partial occultation of the base
of the accretion column, in combination with the angle dependence of the
cyclotron cross section. These physical effects can be studied in an
approximate manner by allowing the parameters in the BW model to vary as
functions of the pulse phase. For example, the effect of the changing
angle of observation due to the pulsar spin can be modeled by allowing
the cross sections $(\sigma_\parallel)$ and $(\sigma_\perp)$ to vary,
resulting in variation of the parameter $\xi$ as a function of the
pulse phase (see eq.~[\ref{eq:firsteq}]). Conversely, the effect of the
partial occultation of the base of the column can be modeled by allowing
for a phase dependency in the parameters $B$, $T_e$, and $\delta$.
Moreover, in each phase bin we note that light bending \citep[see e.g.
][]{kraus2003} might cause the mixing of emission arriving from
different angles. Based on these observations, we extract
phase-dependent parameters by applying the BW model to the spectra in 15
equally spaced phase bins. 

For the BW component, we adopt the values of $\dot M$ and $r_0$
derived for the phase averaged spectrum. For the lower energy
\texttt{compTT} component, since we do not use the \lecs data, the
temperature of the seed photons cannot be constrained; we therefore
freeze it to the value 0.51\,keV found in the phase averaged spectrum.
Depending on the particular characteristics of each single spectrum,
some parameters cannot be constrained and thus are frozen at a suitable
value consistent with the neighboring bins or the phase averaged
spectrum. In particular, we put great care in limiting the line width by
freezing the $\sigma$ parameter in the fitting process whenever its
unconstrained value would give a major contribution to the continuum
model.

\begin{figure}
  \begin{center}
\resizebox{\hsize}{!}{
\includegraphics{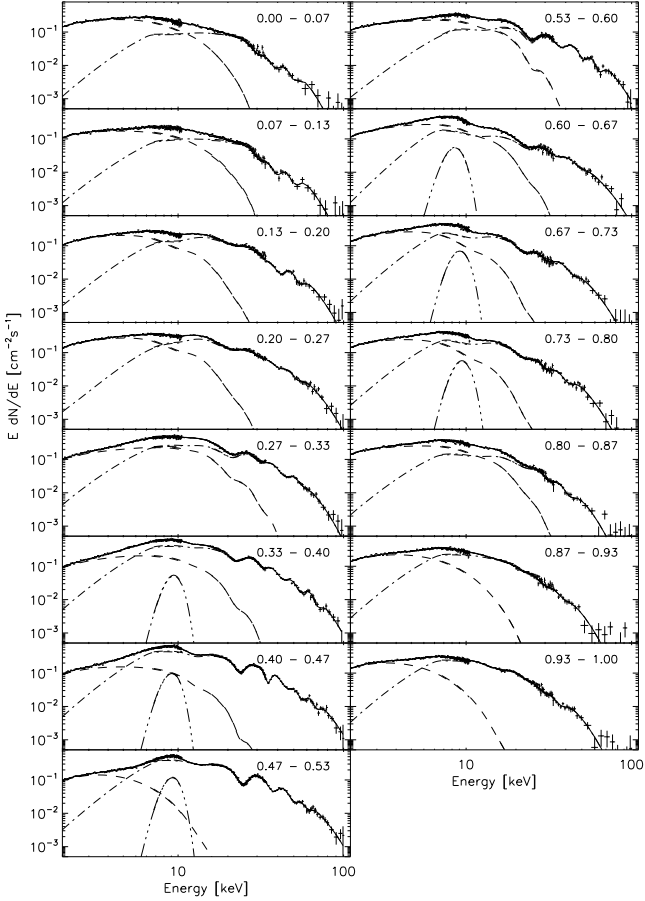}
}
    \caption{Unfolded phase resolved spectra. The phase range is indicated in each plot. The dashed line is the 
		thermal Comptonization, the dot-dashed line the column emission computed using the BW model, the dot-dot-dashed line a narrow Gaussian line, and the solid line the sum.}
\label{fig:spectra}
\end{center}
\end{figure}

We also allow a fine tuning of the relative normalization between the
different instruments due to the uncertainty in the absolute timing for
\sax (see Sect.~\ref{sec:observation}). However, we verified that the
scatter is within 8\% (\pds) and 3\% (\mecs) from the values computed
for the phase averaged spectrum and reported in
Table~\ref{tab:constants}. Eventually, we obtained for the fits reduced
$\chi^2$ between 0.94 and 1.2 for 346--352 degrees of freedom
corresponding to a null hypothesis probability always lower than
0.5\%. Given the model uncertainties, we consider this an acceptable
result. The unfolded spectra are reported in Fig.~\ref{fig:spectra}.

\begin{figure}
  \begin{center}
\resizebox{\hsize}{!}{
\includegraphics[angle=0,width=8cm]{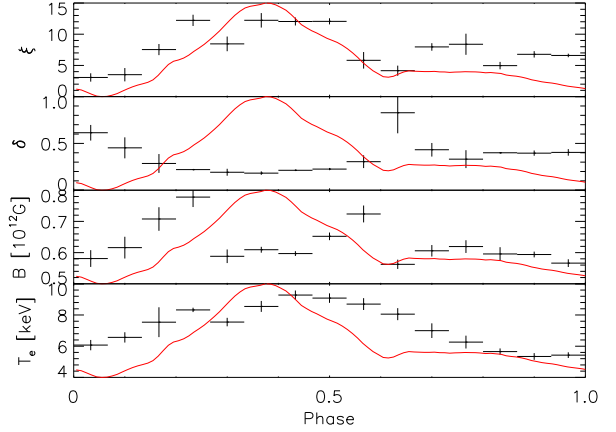}
}
    \caption{Best fit parameters for the Becker and Wolff spectral component
as function of the phase. The accretion column radius is fixed at 600\,m and
the mass accretion rate at $6 \times 10^{16}\,\mathrm{g\,s^{-1}}$. The solid
line is the smoothed pulse profile in the 10--20\,keV energy band with arbitrary
normalization.
}
\label{fig:BW}
\end{center}
\end{figure}

The phase dependency of the best fit parameters in the BW component is
shown in Fig.~\ref{fig:BW}: the magnetic field $B$ is higher at the
edges of the main peak (phase 0.13--0.66), as is the contribution of
bulk with respect to thermal Comptonization described by the parameter
$\delta$. The parameter $\xi$ follows roughly the luminosity phase
dependency, whilst the temperature of the electrons $T_e$ displays a
single peak at phase $\sim 0.5$. The pattern is less clear outside the
main peak and will not be discussed further.

\begin{figure}
  \begin{center}
\resizebox{\hsize}{!}{
\includegraphics[angle=0,width=8cm]{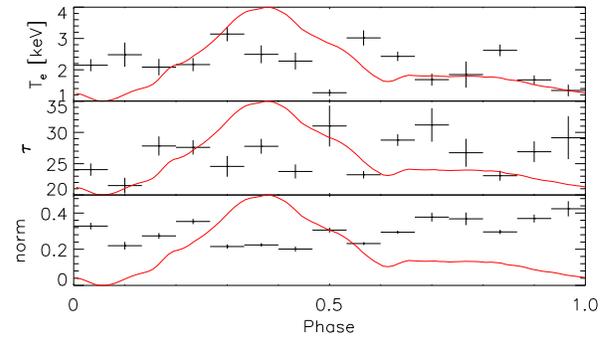}
}
    \caption{Best fit parameters for the soft Comptonization spectral component (compTT) as function of the phase.  The temperature of the seed photons is fixed to 0.51\,keV. The solid line is the smoothed pulse profile in the 10--20\,keV energy band with arbitrary normalization.
}
\label{fig:compST}
\end{center}
\end{figure}

\begin{figure}
  \begin{center}
\resizebox{\hsize}{!}{
\includegraphics[angle=0,width=8cm]{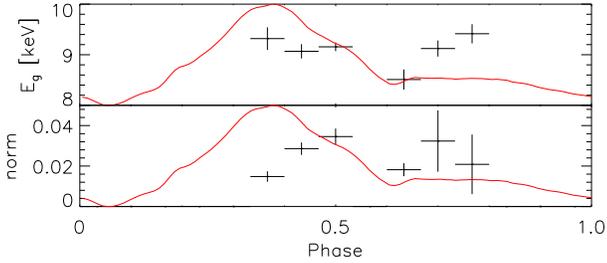}
}
    \caption{Best fit parameters for the Gaussian emission line as function of the phase. The line width is fixed to 1\,keV. The solid line is the smoothed pulse profile in the 10--20\,keV energy band with arbitrary normalization.
}
\label{fig:Gaussian}
\end{center}
\end{figure}

The phase dependency of the optical depth $\tau$ and electron
temperature $T_C$ in the lower energy \texttt{compTT} component is shown
in Fig.~\ref{fig:compST}: the former is always very high, between 20 and
35, and the latter varies almost by a factor of three. Even though the
deviation from a constant value is highly significant for both
parameters, there are no clear correlations with the pulse structure.
The additive Gaussian line has always a centroid energy around 9\,keV
with no obvious correlation to the magnetic field intensity
(Fig.~\ref{fig:Gaussian}), however its minimal presence is required to
obtain an acceptable $\chi^2$ in six phase bins.

\begin{figure*}
  \begin{center}
	\includegraphics[angle=0,width=5cm]{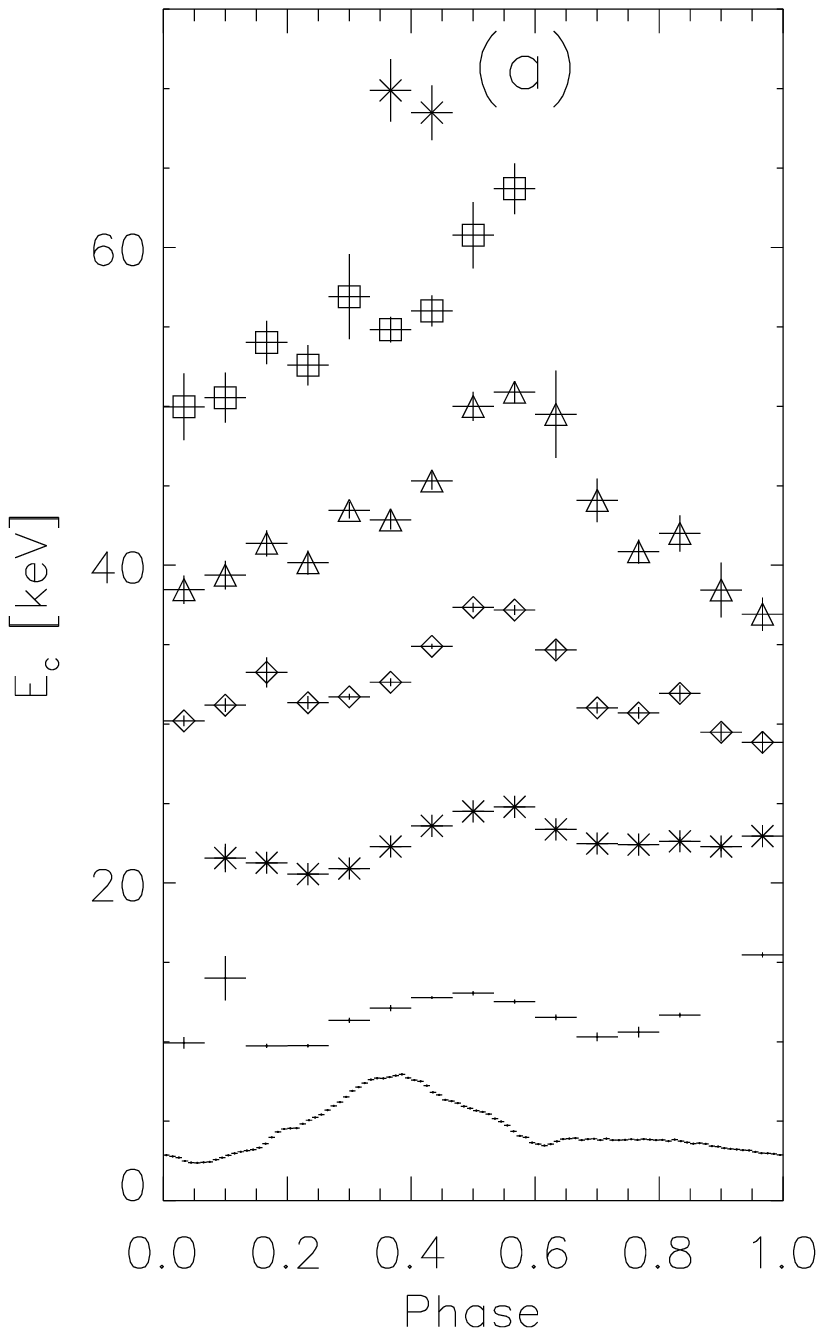}
	\includegraphics[angle=0,width=5cm]{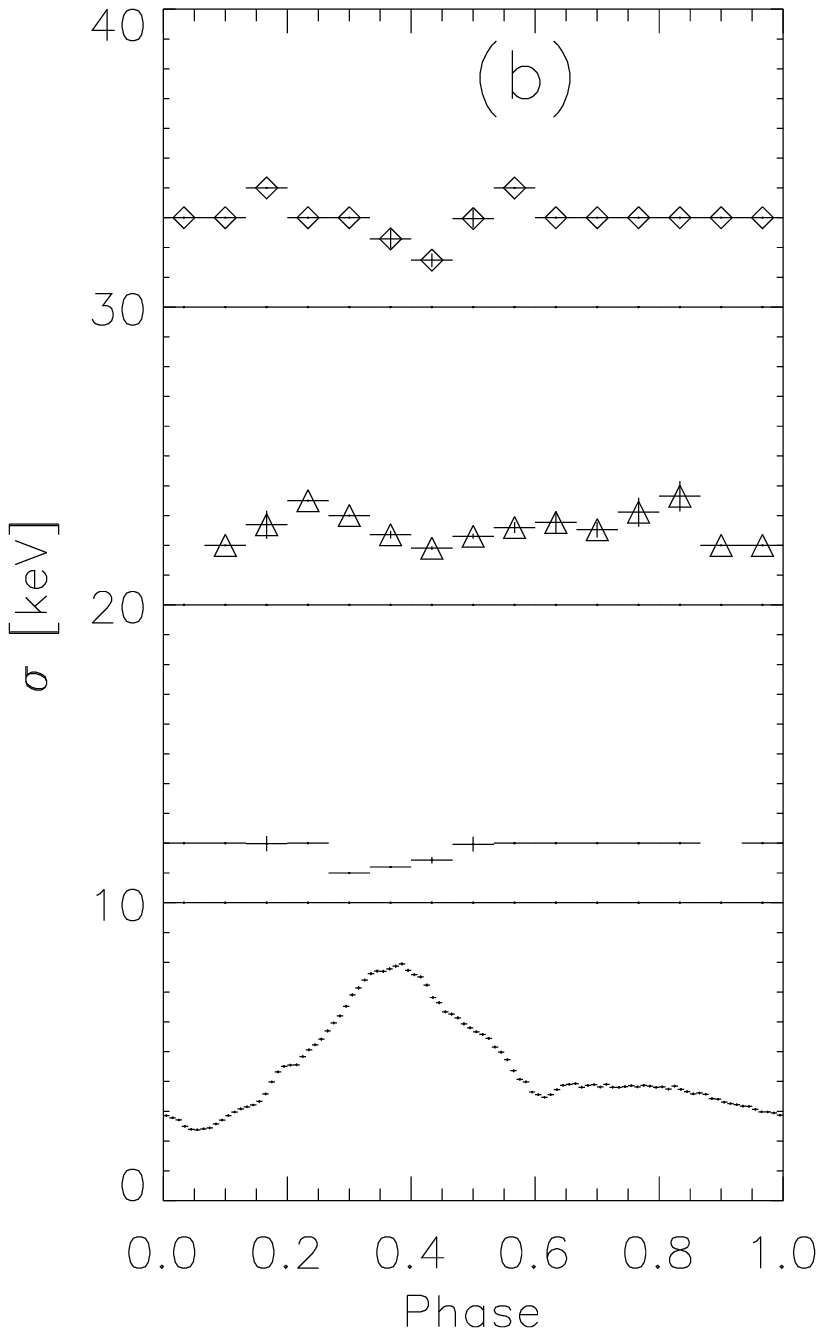}\\
     \includegraphics[angle=0,width=5cm]{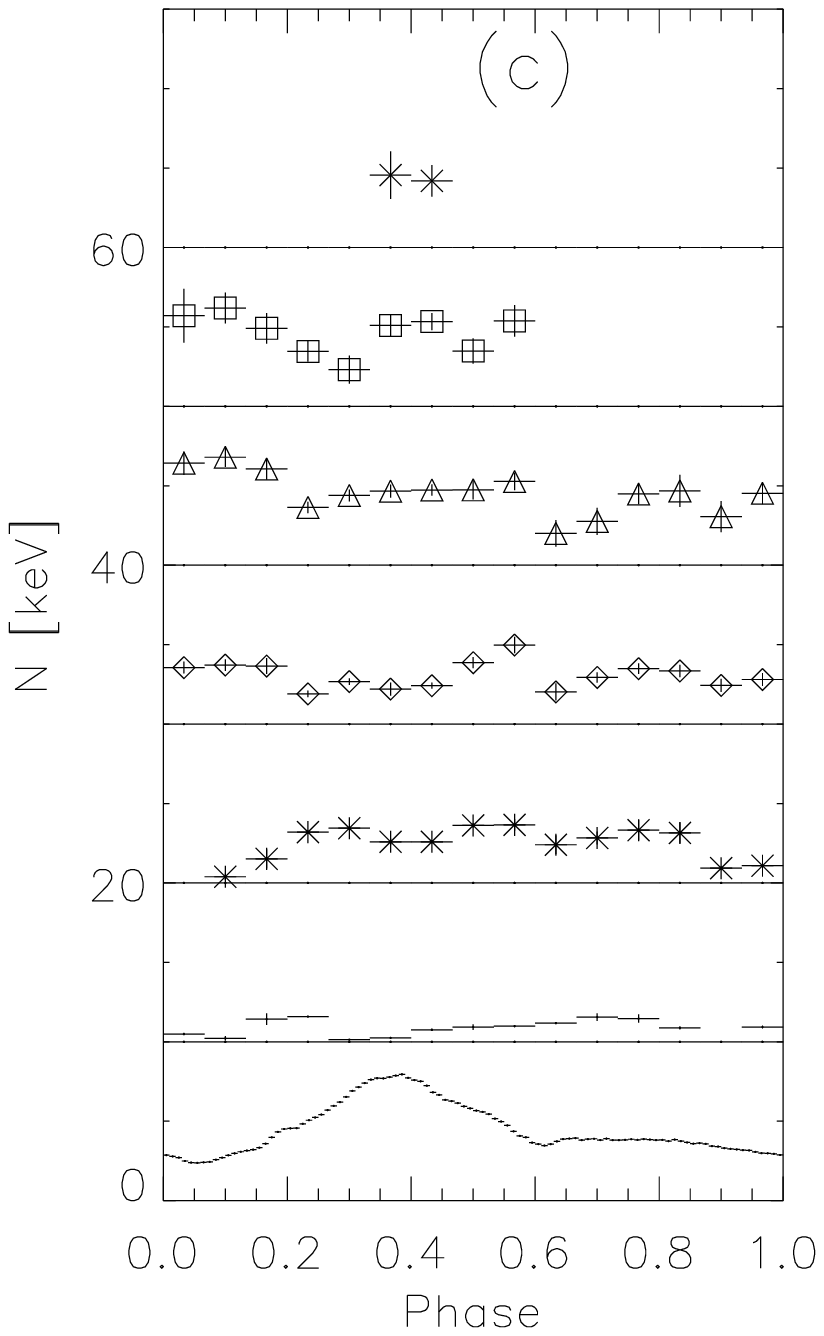}
     \includegraphics[angle=0,width=5cm]{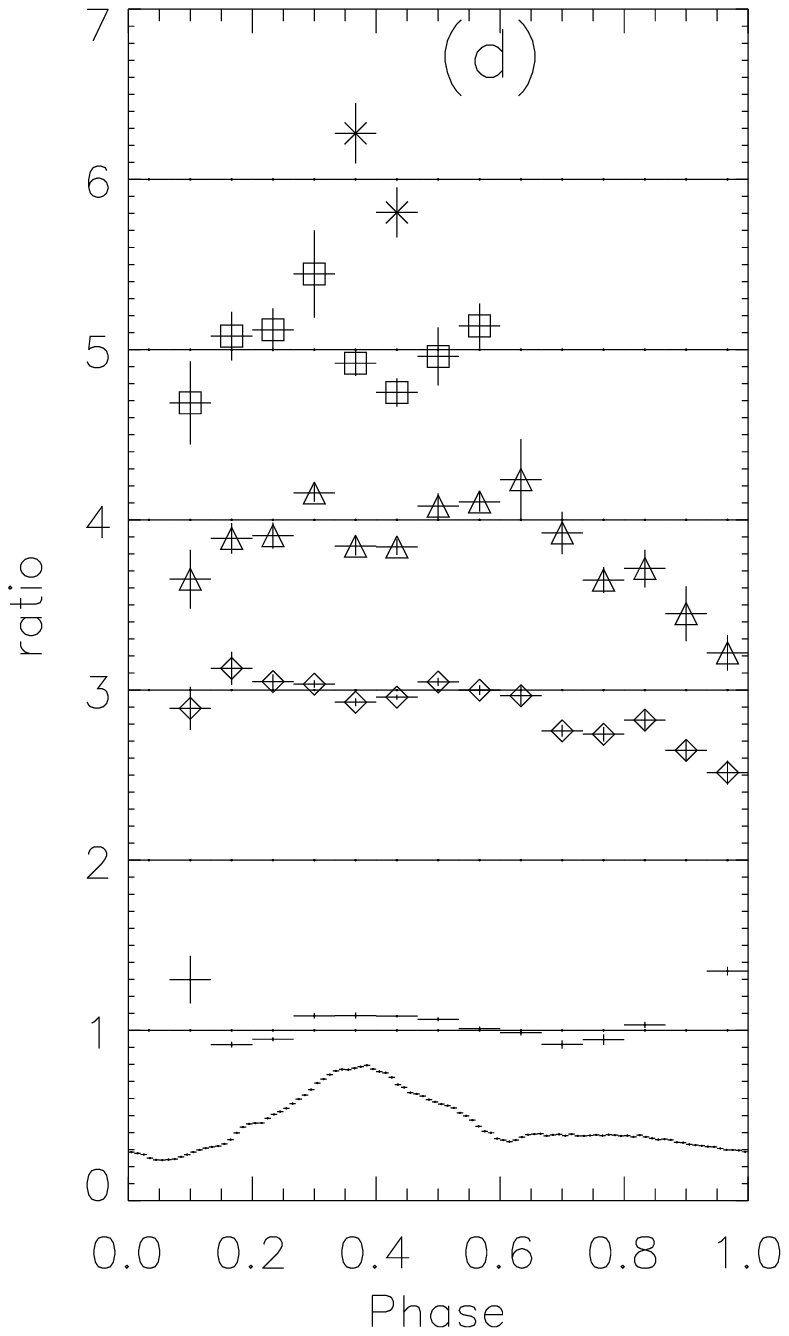}
    \caption{Absorption line parameters. (a) The centroid energy. (b) The width. (c) The normalization. (d) The ratio between the centroid energy and half the 
	     centroid energy of the second harmonic. In each plot the 10-20\,keV pulse profile is sketched with arbitrary normalization for illustrative purpose.
	     Line width and normalization are vertically displaced of $10 n$\,keV, where $n$ is the harmonic order. Uncertainties are reported at 
	     $1\sigma$ c.l. assuming a Gaussian symmetric distribution.
    }
\label{fig:lines}
\end{center}
\end{figure*}

We show in Fig.~\ref{fig:lines} the centroid energies, the width of the
first three harmonics, the line normalization, and the ratio between the
centroid energies and half the centroid energy of the second harmonic.
This final parameter can be considered the best indicator of the magnetic
field intensity in the line forming region, since the second harmonic is
purely due to scattering and, incidentally, it is covered with high S/N
by two instruments in this observation. We note that along the main peak
(phase 0.12--0.66), the higher harmonics are almost equally spaced: the
deviations are at most of 4\% with less than 3$\sigma$ significance.
Note that for the same phase range, the fundamental is systematically higher,
and from phase 0.66 to 1.12, the lines do not present a regular spacing.

\subsection{Search for higher harmonics}

At phase 0.33--0.46 (the peak), we introduce a previously undetected
{\it sixth harmonic} to model some residuals in the high-energy tail of
the spectrum. To have a better grasp on the line significance and verify
it is not due to an inappropriate modeling of the continuum, we test its
presence using a cut-off power model for the energy range above 13\,keV.
This model was used for previous analyses \citep[see e.g.
][]{santangelo1999}. The results are reported in Table~\ref{tab:sixth}:
the probability that the feature is only a statistical fluctuation is
not negligible, therefore we cannot claim a firm detection of the sixth
harmonic, just a strong suggestion of its presence since the centroid
energy is compatible with the value of the second harmonic, the line
normalization is greater than zero at 90\% c.l., and it is present in
two distinct phase bins at the pulse maximum.

\begin{table*}
\begin{center}
\caption{Fit results in the search for the $6^\text{th}$ harmonic using
the cut-off power law model for the continuum above 13\,keV.
$\Delta\chi^2$ is referred to the introduction of the further Gaussian
absorption line. $E_\text{abs}$ is the centroid value, $E_B$ is the
value corresponding to three times the centroid energy of the second
harmonic, and $N\text{abs}$ is the line normalization. The null
hypothesis probability is obtained using the F-test. 
}
\begin{tabular}{l c c}
\hline
\hline
phase                        & 0.33--0.40                  & 0.40--0.46              \\
\hline
$E_\text{abs}$ [keV]         & $71\pm3 $                   & $68.6\pm2.6$              \\
$E_B$ [keV]                  & $66.9\pm0.3$                & $70.74\pm0.15$             \\
$N_\text{abs}$ [keV]         & $4.5\pm2.5$                 & $3.1_{-0.6}^{+1.3}$         \\
$\Delta\chi^2\text{(d.o.f.)}$&$221(168)\rightarrow215(166)$&$158(142)\rightarrow152(140)$ \\
Null hypothesis probability  &  10\%                       & 5\%                           \\
\hline
\end{tabular}
\label{tab:sixth}
\end{center}
\end{table*}

%

\section{Discussion}
\label{sec:discussion}

We have applied the self-consistent bulk+thermal Comptonization model
developed by \citet{becker2007} to study the bright transient source
\src. In order to treat this source, we find that the BW model needs
some adjustments with respect to the prescriptions in the original
paper. The first issue is the relatively low accretion rate compared to
the X-ray luminosity: assuming a distance of 7\,kpc, we find a
1-100\,keV luminosity of $6.3\times10^{37}\,\mathrm{erg\,s^{-1}}$ which
corresponds to $\dot M=3.3\times 10^{17} \,\mathrm{g\,s^{-1}}$ for
unitary conversion efficiency from gravitational to electromagnetic
energy. If we consider only the BW component, we find $L_X = 5.5 \times
10^{37}\,\mathrm{erg\,s^{-1}}$ and $\dot M=1.7\times 10^{17}
\,\mathrm{g\,s^{-1}}$. Our adopted value for $\dot M$ is instead
$0.6\times 10^{17} \,\mathrm{g\,s^{-1}}$, a factor three lower. However,
using the relation $L_X=G M_* \dot M/R_*$, one assumes an isotropic
emission over $4 \pi$ steradians. This clearly overestimates the
luminosity from a single magnetic pole by at least a factor of two. The
factor could be even larger if the emission came from a fan beam
geometry and/or the NS has a different mass to radius ratio. In general,
the prescription for the calculation of the luminosity in X-ray pulsars
probably needs to be re-examined and firmed up due to these expected
anisotropies in the emission profile.

The most striking observational feature of \src is the unique presence
of at least five cyclotron lines in its spectrum; other HMXBs display at
most a second, and in one case a third harmonic \citep[see
][]{coburn2002,orlandini2004,tsygankov2006,isabel2007}. This is due to
the relatively low intensity of its magnetic field, the lowest among the
Galactic HMXBs that display cyclotron absorption features. In fact, at
energies corresponding to the higher harmonics, there is still a high
photon density, up-scattered in energy by the thermal population of
electrons at a temperature of a few keV. The presence of the significant
high-energy continuum allows the harmonic absorption features to be
detected.

This peculiar characteristic finds some correspondence also in the
continuum model. For example, we show for the first time that the high
energy ($\ga 10$\,keV) emission of \src is dominated by Comptonized
cyclotron cooling, as expected on theoretical basis \citep{arons1987}.
Conversely, in Her~X-1 and other X-ray pulsars with higher magnetic
field, the emission is dominated by the Comptonized bremsstrahlung
component \citep{becker2007}. Using the best fit parameters obtained
from the phase averaged spectrum, we simulated the source emission if
the magnetic field were higher and found that the relative contribution
of the cyclotron component decreases relative to bremsstrahlung and
fades to a negligible value for $B \ga 5\times 10^{12}$\,G. The reason
is the decreased production rate for the cyclotron seed photons that
form as a result of radiative de-excitation from the first excited
Landau level, since this is less and less populated as its energy
becomes higher than the electron temperature. As a consequence, the
spectrum of this source is dominated by reprocessed cyclotron emission,
in contrast to other bright pulsars with stronger magnetic fields, in
which the emission is dominated by reprocessed bremsstrahlung emission
\citep{becker2007}.


\begin{figure}
  \begin{center}
\resizebox{\hsize}{!}{\includegraphics[angle=0]{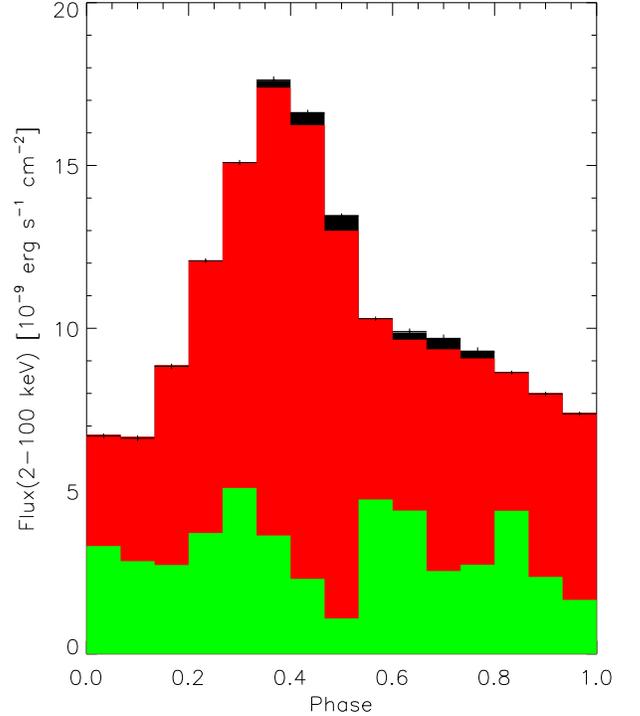}}
         \caption{Contribution of the BW model plotted in red (dark gray), compTT in green (light gray), and the Gaussian emission line in black to the total flux (points with error bars). }
\label{fig:flux_deconvolution}
\end{center}
\end{figure}

Decomposing the contributions of the different components as functions of
the phase demonstrates that along the peak, between 65\% and 90\% of the
radiation is produced by the BW component, while at other phases the
relative contribution of the thermal Comptonization reaches $\sim$50\%
(Fig.~\ref{fig:flux_deconvolution}). The contribution of the Gaussian
emission line is always less than 5\% and is concentrated on the
descending part of the main peak. This is probably due to incomplete
modeling of the cyclotron scattering process. The high energy
component clearly originates in the accretion column while the low
energy component has the characteristics of a diffuse halo, since
its absolute flux is roughly constant throughout the phase.

\subsection{Parameter constraints}

During the determination of the optimal mass accretion rate, we noted
that this parameter is tightly anti-correlated with the column radius,
and both are linked to $\xi$ via eq.~(\ref{eq:firsteq}). Since there is
no basis for choosing a priori the value of $\xi$, it was necessary to
study the probability contour plots of $r_0$ and $\dot M$ in the range
$450\,\mathrm{m} < r_0 < 1100\,\mathrm{m}$ and $0.4 < \dot M_{17} <
1.1$, where $\dot M_{17}$ is the mass accretion rate in units of
$10^{17}$\,g/s. We found that at 68\% c.l., any combination of these
parameters is acceptable provided it falls between the lines of the
equations $r_0 = 1250\,(\dot M_{17} - 0.56) + 450$ and $r_0 = 1823\,
(M_{17} - 0.46) + 450$, where $r_0$ is expressed in meters. We then made
the choice of fixing $\dot M_{17}=0.6$ and $r_0=600$\,m, the latter
according to the indication of a relatively large column in other bright
sources studied by \citet{becker2007}, and also based on the upper limit
set by \citet{lamb1973}, and later revised by \citet{harding1984}.

We also verified that the NS mass and radius cannot be constrained by
the spectral data. Any change in their value within reasonable
boundaries can be compensated by tuning the mass accretion rate and
column radius. Therefore, only four out of the namely six free
parameters can be determined by a direct fit of the model to the data.
On the contrary, we already showed that we can constrain just a
relatively wide region of the $r_0$--$\dot M$ plane, and this region
depends on the particular choice of the NS base parameters. Among the
free parameters, the magnetic field is insensitive to the choice of the
fixed parameters, while $\xi$, $\delta$ and $T_e$ depend on them. At
least in the case of \src, where the lower energy portion of the
spectrum is dominated by another component, it is not possible to infer
from the data all the physical characteristics of the accretion column.
Nevertheless, the BW model gives us the unique chance to investigate the
basic emission mechanisms.

\subsection{Magnetic field structure and emission geometry}

The other main issue is the strength of the magnetic field producing
the cyclotron emission, which is found to be $(0.60\pm0.06)\times
10^{12}$\,G, while in the line forming region the field strength is
$0.989\times 10^{12}$\,G, as derived from the centroid energy of the
second harmonic (the values are given neglecting the gravitational
redshift, which acts as a scale factor). Assuming a dipolar field, and
supposing that the cyclotron absorption features form close to the
neutron star surface, we derive the cyclotron emission is concentrated
at $\sim$1.7\,km above the stellar surface. Based on this observation,
we argue that the cyclotron cooling in \src takes place between the NS
and the column sonic point, which is at $\sim$2.7\,km above the surface,
according to the best fit model parameters. At that height, the emission
is strongly beamed since the plasma speed is $v/c\sim 0.25$ and most of
the radiation is emitted downwards, as predicted for an emitting charge
in relativistic motion \citep[see e.g. eq. 4.100 -- 4.103 of
][]{rybicki1979}.

The radiation is subject to Compton reprocessing inside the accretion
column, and it can subsequently escape from the sides only near the
base, where the advection velocity decreases, and the density and
temperature are higher. We can verify that for the best fit value of
$\xi$, the diffusive escape time from the column is comparable to the
free-fall time from the sonic point -- see eq.~(104) of
\citet{becker2007}. Further confirmation of this scenario comes from the
phase dependency of the parameters $\xi$ and $T_e$, which are higher
along the peak, implying respectively that the radiation traveled mostly
parallel to the magnetic field, and that the visible layer is deeper in
the column for this phase range. The sharp feature in the pulsed
fraction (Fig.~\ref{fig:pulsed_fraction}) at $\sim20$\,keV also fits
into the picture: while the emerging radiation is beamed roughly
perpendicular to the column, the resonant scattering produces an
isotropic emission which reduces locally the pulsed fraction.

The lower energy component is due to thermal Comptonization of seed
photons with temperature $T_0=(0.51\pm0.01)$\,keV, which is not far from
the temperature of the thermal mound probably present at the base of the
column: according to the BW model of the high energy emission we find
$T_{\mathrm{Th}}\sim0.7$\,keV. However, any attempt to fit the broad
band spectrum using the BW model alone was unsuccessful due to the very
low flux of the components originating from the blackbody and
bremsstrahlung sources compared to the cyclotron one. As was already
discussed, the effective radius of the blackbody photon source for the
low energy \texttt{compTT} component is about 15\,km (or 9\,km
considering the gravitational redshift). It is therefore likely that the
low-energy radiation source resides close to the NS surface, perhaps in
a diffuse halo. An optically thick atmosphere $(\tau \sim 20)$ at a
temperature of a few keV could be confined by multipolar or crustal
components of the magnetic field \citep[see e.g. ][]{gil2002}, while the
region around the column could be heated to $\sim0.5$\,keV either by the
accretion itself or by the strongly beamed column emission.

The modulation of the magnetic field producing the continuum radiation
along the main peak (Fig.~\ref{fig:BW}) suggests that the emission layer
has not a planar geometry, but becomes closer to the NS at the edges of
the column. A similar pattern is present in the phase dependency of the
$\delta$ parameter, indicating that bulk Comptonization is stronger at
the center of the main peak. The photon-diffusion parameter, $\xi$,
follows nicely the pulse profile in the phase interval 0.2-0.8. This can
give us some information on the angle between the magnetic field and the
radiation direction.

\begin{figure}
  \begin{center}
     \resizebox{\hsize}{!}{\includegraphics[angle=0]{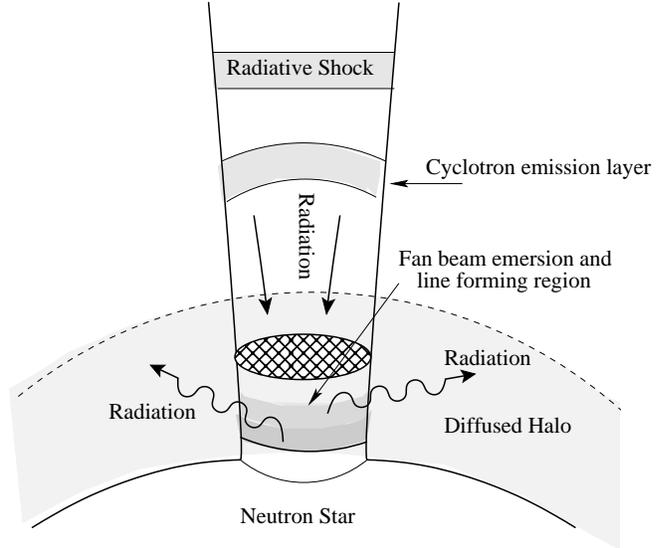}}
         \caption{A schematic and not in scale cartoon of the magnetospheric structure of \src,
         	 as it emerges from our analysis.}
\label{fig:cartoon}
\end{center}
\end{figure}

Even though $\xi$ is inversely proportional to the geometrical mean of
the extreme values of the electron scattering cross section (see
eq.~[\ref{eq:firsteq}]), and does not include explicitly its angle
dependence, it can be regarded as an estimator of the average value in a
particular phase bin. We argue that at the maximum of the peak, the
radiation field is subject to a reduced scattering cross section, and
therefore the photons must travel preferentially in the direction of the
magnetic field lines \citep{ventura1979}. The photons are therefore able
to reach almost the NS surface, before escaping from the column walls
mostly near the base, where the plasma temperature is higher. At the
column boundaries, where the plasma is optically thin outside the
cyclotron resonances, the observed cyclotron absorption features are
imprinted on the spectral energy distribution. In Fig.~\ref{fig:cartoon}
we provide a schematic drawing of the model that we propose.

\subsection{Standard spectral phenomenological model}

Even though we were able to successfully model the continuum with a
standard spectral phenomenological model similar to the one that \citet{dima2008}
used for EXO~2030+375, we consider this model unsatisfactory for \src
due to its incoherency with the pulse energy dependency. The pulse
profiles and the pulsed fraction change at $\sim 7$\,keV, corresponding
to the spectral region where the Gaussian ``bump'' starts to dominate,
whilst they do not exhibit a peculiar evolution at the energy where the
power law continuum becomes again the main component. It is also
difficult to explain why the same spectral component produces a pulsed
fraction which is nearly constant below $\sim5$\,keV and steadily
increasing above $\sim15$\,keV.

Adopting the spectral phenomenological model, we note the appearance of some
residuals in the spectral fit below $\sim1$\,keV
(Fig.~\ref{fig:phase_averaged_pheno}). Even though they are not
statistically significant due to the low exposure of the \lecs
instrument, they essentially disappear when we adopt the multi-component
Comptonization model (Fig.~\ref{fig:phase_averaged}). With this latter
model, we find that the photoelectric absorption due to neutral plasma
is entirely due to the foreground Galactic medium: this is plausible
since the observation took place after the outburst maximum when the
strong radiation field of the NS could have ionized the surrounding
medium.

Moreover, to our knowledge there is no physical process which could
explain the broad Gaussian ``bump'' in emission: for instance, if it is
due to cyclotron cooling, then one must find a way to make the Compton
scattering so ineffective in reprocessing this component that it
maintains a broad Gaussian profile instead of the typical power-law with
exponential decay. We conclude that although this model can yield a good
phenomenological description of the spectrum, it is nonetheless unable
to give a coherent picture of the system in terms of physical processes.

The decoupling between the high and low energy emission is more
naturally explained by two distinct emission components, one dominating
below $\sim7$\,keV, and the other above. In the scenario we have focused
on here, the low energy emission (48\% of the phase averaged total flux
in the 1--100\,keV energy band) is modeled as a thermal Comptonization
of 0.51\,keV seed photons \citep{comptt}, while the high energy
continuum (51\% of the total flux) is modeled using the BW formalism
\citep{becker2007}. We also add a Gaussian emission line which gives a
minor contribution to the total flux ($\sim$~1\%). The appearance of
this component likely reflects either an approximation of the emission
wings predicted to appear near the fundamental by Monte Carlo
simulations of cyclotron scattering \citep[][ and references
therein]{gabi2007}, and/or an adjustment to the BW model to account for
a more complex geometry in the scattering process. While this component
is required to obtain an acceptable $\chi^2$, it is clearly not a major
contributor to the continuum, in contrast to the phenomenological model
described above, in which it contributes 16\% of the total flux in the
1--100\,keV energy band.

\subsection{Structure of the cyclotron harmonics}

The line widths of the second and third harmonics reach a
minimum just after the maximum of the pulse, which indicates that
the line of sight reaches its maximum angle relative to the magnetic
field at this phase. However, we note that the lines appear much
shallower than those computed in the MC models \citep{gabi2007}.
Probable reasons are a more complex geometry with respect to the model
assumptions and general relativistic effects such as light bending, lensing
effect, and doppler shift. A more
quantitative discussion of this issue and the scenario we envisage
for the system would require a complete numerical simulation, but this goes
beyond the scope of the present paper.

If the radiation field dominates the thermodynamics of the plasma,
then one expects a link between the temperature of the plasma and the
magnetic field intensity. The relation depends on the optical depth in
the absorption lines, the angle that the radiation forms with the
magnetic field lines, and the system geometry \citep[e.g. ][ and
references therein]{isenberg1998}. In a slab geometry and for radiation
injected parallel to the magnetic field, the cyclotron energy should be
more than two times the plasma temperature. For isotropic radiation, one
instead expects a ratio of about 4. This is the case for the halo
temperature and the cyclotron line energy found here, while the column
temperature is too high to be due to cyclotron heating, supporting the
idea that the line forming region is not located within the column but
rather in an external halo.

The apparent non-harmonicity of the line spacing in some HMXBs has been
explained by the variation of the magnetic field in the line forming
region. \citet{nishimura2003} showed that for a dipolar field the height
of the region relevant for the absorption and scattering process must be
of the order of several hundred meters to reproduce the observations. In
a later work \citet{nishimura2005} demonstrates that a linearly varying
magnetic field is able to produce line non-harmonicity even for a 5\,m
thick slab of plasma above the NS. Such a configuration can be modeled
by a combination of axial and crustal magnetic field, as proposed by
\citet{gil2002}.

We show that the higher harmonics are equally spaced in \src, as also
noted by \citet{santangelo1999} and \citet{russi2007}, and therefore the
field is nearly constant in the line forming region. The previously
reported deviations from harmonicity regards only the ratio to the
fundamental, whose centroid energy can be displaced from the cyclotron
value. This can be due to the influence of the other emitting component,
which is present below the cyclotron energy, and/or to a complex shape
of the line profile, heavily modified by photon spawning, and then
poorly described by a Gaussian absorption line. Also the need to
introduce a Gaussian emission line points towards a complex structure
of the cyclotron fundamental line, with emission wings that are not
modeled by a Gaussian profile.

Using the second harmonic, with energy $\sim20$\,keV, as the best
estimator of the magnetic field in the line forming region, the
displacement from the harmonic law is always less than 5\% for the
higher harmonics, thus the scattering region thickness is at most 2\% of
its distance from the NS center. The modulation of the line centroid
energy with phase was already reported e.g. by \citet{mihara2004} and is
a typical feature of this source: it is interpreted as due to the
occultation of part of the column by the NS along its rotation. Since
the magnetic field of the line forming region varies from $(1.07 \pm
0.08)\times 10^{12}$\,G to $(0.87 \pm 0.15)\times 10^{12}$\,G along the
main peak, the maximum reduction from the maximum value is 23\%. This
corresponds to an 8\% variation of the height of the line forming
region; assuming that at phase 0.5--0.6 the absorption takes place close
to the base of the column and the NS has a radius of 10\,km it means
that for instance at phase 0.2--0.3 the scattering takes place in a
layer, which is a few hundred meters thick, located $\sim 800$\,m above
the NS surface.

The last issue we would like to briefly discuss is the puzzling jump of
the energy of the fundamental cyclotron feature around the pulse minimum
from $\sim11$ to $\sim15$\,keV. A similar discontinuity was already
noted in the phase averaged spectrum by \citet{mihara2004,nakajima2006,
russi2007} when the luminosity drops below $5\times10^{37}$\,erg/s: it
is interpreted as a drift of the compact emitting region in the magnetic
field, due either to the disruption of the radiative shock or possibly
to the accretion column burning out with the onset of another column in
a multi-polar configuration. We show that this jump is not linked
exclusively to the system luminosity, but appears at different phases
even when the source is brighter than the proposed threshold for the
shock disruption ($\sim6\times10^{37}\,\mathrm{erg\,s^{-1}}$). Moreover,
the absorption features at 30--40\,keV which are present outside the
main peak, do not seem to be linked by an harmonic ratio to the
$\sim20$\,keV line, which suggests that complex field geometries could
be present in \src.

\section{Conclusions}
\label{sec:conclusions}

For the first time we apply the self-consistent model of the high
luminosity X-ray binary pulsar emission proposed by \citet{becker2007}
to the well studied transient source \src during the giant outburst that
occurred in 1999. We find that the emission of the peak is produced by
thermal and bulk Comptonization of the cyclotron emission which is the
main cooling channel of the column. On the other hand the lower
energy continuum is roughly constant throughout the phase, and is due to
thermal Comptonization of a 0.5\,keV blackbody, whose effective
radius is comparable to the NS dimensions.

We are able to sketch a model in which the photons are generated by
cyclotron cooling well above the NS surface. The photons are advected
towards the NS by the relativistic plasma bulk motion, while they are
Compton up-scattered by the hot electrons in the column. The radiation
escapes from the column lateral walls close to the NS surface, where
the relativistic beaming is less severe. At this height the cyclotron
lines are formed, probably in the boundary layer between the column and
the plasma which forms an extended halo above the NS surface, and also
produces the low-energy spectral component.

Unequivocal interpretation of the phase-dependent parameters using the
theoretical model would require some estimate of the angles for the
system. I.e., the angle between the line of sight and the rotation axis,
and the angle between the rotation axis and the accretion column. These
two angles would determine the portion of the column visible during a
given phase, as well as the angle between the normal to the column and
the line of site to the observer. Relativistic ray tracing should be
used to study the influence of the system geometry on the emission
pattern, but this is beyond the scope of this paper.

Although we have developed a convincing physical model for \src which
is able to describe coherently the system properties, we emphasize that
this scenario should be checked quantitatively in future work using a
detailed simulation. The analysis presented here was made possible via
the implementation of the model proposed by \citet{becker2007} in the
standard fitting package \XSPECB. The application of this new
model is very promising and it is expected to facilitate the study of
many other systems.

\section*{Acknowledgements}
The authors would like to thank M. Wolff for suggesting to us the
possible anisotropy of the X-ray emission from the column, and
also the referee M. Falanga who provided several insightful comments
that led to significant improvements in the manuscript. C. F. has been
supported by grant DLR~50~OG~0601 during this work.

\bibliographystyle{aa}

\end{document}